# Toward an usage of americium-241 for filter calibration

G. Dougniaux[1], B. Sabot[2], S. Pierre[2], B. Dhieux Lestaevel[1]

[1] Institut de Radioprotection et de Sûreté Nucléaire (IRSN), PSN-RES/SCA/LPMA, F-91400 Saclay, France

[2] Université Paris Saclay, CEA, LIST, Laboratoire National Henri Becquerel (LNE-LNHB), F-91120 Palaiseau, France

## Abstract

In nuclear safety and environmental studies, it is necessary to measure the activity of aerosols collected on filters. The calibration of these filters' measurement chains currently involves producing a reference filter with a controlled deposit of aerosol marked by a specific radionuclide. In France, commonly used radionuclides include $^{137}$Cs for gamma or beta emitters, $^{90}$Sr/$^{90}$Y for beta emitters, and $^{239}$Pu for alpha emitters.

However, using $^{239}$Pu presents several issues, notably the destruction of the filter required to determine its traceable activity and the associated uncertainties. The challenge of this project is to propose a radionuclide enabling non-destructive, SI-traceable measurement while retaining the reference filter. The proposed radionuclide is $^{241}$Am, which is both alpha and gamma-emitting and allows for precise non-destructive measurements.

Reference filters are produced on the IRSN's ICARE test bench with calibrated aerosols tagged with $^{241}$Am. The reference activity of theses filters is measured by gamma spectrometry on CEA/LNHB reference counter.

A series of tests confirmed that $^{241}$Am provides accurate activity determination with minimal uncertainty. Consequently, adopting $^{241}$Am can significantly improve the reliability of radioactive contamination monitoring while avoiding practical challenges associated with plutonium.

Key-words: americium, filter, calibration, aerosol.

## 1. Introduction

In the field of nuclear safety and environmental studies, it is essential to measure the radioactive activity of aerosols collected on filters. Organisations that are required to monitor airborne radioactive contamination within facilities or in the environment collect air samples on filters and subsequently measure the radioactivity of these filters. The results are then conveyed in the form of reports on the prevailing state of radioactive contamination in the atmosphere. With regard to environmental radioactivity measurements, these organisations constitute a national network for environmental radioactivity monitoring, established by articles R. 1333-25 and R. 1333-26 of the French Public Health Code. This network contributes to the estimation of the doses of ionising radiation to which the population is exposed and provides public information. The national network has two principal objectives:

- Strengthening transparency regarding information about environmental radioactivity;
- Developing a quality policy for environmental radioactivity measurement results.

The organisations in this national network have been accredited by the ASN (*Autorité de Sûreté Nucléaire*) on the basis of their capacity to provide accurate and reliable measurements of atmospheric contamination. These accreditations are classified into several categories, each corresponding to a different measurement matrix, including water, soil, biological materials, aerosols, and gases. The primary challenge lies in the agreement 4_03, which pertains to gross alpha on filter. Twenty organisations are accredited for this measurement as of July 2024 (asn.fr). Laboratories engaged in the measurement of airborne radioactivity face significant challenges in calibrating their instruments. To obtain approval, laboratories must provide both the emerging alpha and the true alpha activities, that is to say, the radiation level observed by the detector and the real filter activity. The calibration of measurement systems currently involves the creation of a reference filter, which contains a deposit of aerosols marked by a specific radionuclide. In France, the most commonly used radionuclides include $^{137}$Cs for gamma or beta emitters, $^{90}$Sr/$^{90}$Y for beta emitters, and $^{239}$Pu for alpha emitters. Laboratories generally lack the necessary radiochemistry, or indeed a standard, as $^{239}$Pu can only be measured after complete destruction.

The calibration of a filter is a complex process that is conducted in collaboration with several specialised laboratories. Nevertheless, the utilisation of $^{239}$Pu is associated with several disadvantages, most notably the necessity to destroy the filter to ascertain its activity in a verifiable manner, which introduces an element of uncertainty. In order to ascertain the activity of $^{239}$Pu, a national metrology institut, such as CEA/LNHB in France (lnhb.fr), carries out a complete destruction of the filter. After the incineration, an acid digestion is employed to solubilise any residual materials. The sample is measured using a low-background liquid scintillation counter. Subsequently, a radioactive tracer of the same nature as the filter's radionuclide is introduced into the sample (typically a drop of solution with a few kBq of activity) – standard addition method. The modified sample is measured using the primary Triple-to-Double Coincidence Ratio (TDCR) method (Broda, 2003; Cassette et Vatin, 1992). This provides the reference activity of the modified sample, and a post-measurement calibration is conducted using the low-background scintillation counter. This yields the activity present on the filter, but it results in the destruction of the original filter and makes it impossible to perform further measurements on the same sample.

To collect aerosols from the environment and measure the potential radioactivity they carry, air is filtered through a medium, capturing the aerosols. Various filter sizes and sampling speeds are used in this process. Within this measurement network, three common filter sizes are noted: 130/110 mm, 51/47 mm, and 47/37 mm (total diameter/active diameter). The first dimension refers to the total diameter of the filter, while the second represents the sampling diameter. Airflow rates during sampling range from 0.5 m·s$^{-1}$ to 2 m·s$^{-1}$. The filter material typically consists of a fibrous mixture of cellulose and glass fibres in varying proportions. Choosing the appropriate filter involves balancing efficiency (as high as possible), pressure loss (as low as possible), and cost. Sampling duration can vary, sometimes lasting up to a week, after which the filter is retrieved and analysed.

Filter activity is given in terms of gross alpha, i.e. the total alpha count from the filter, expressed in terms of activity from a reference radionuclide. Gross alpha is thus an indicator of the activity deposited on the filter. This initial measurement is typically conducted before any more precise and complementary analysis is performed if significant activity is detected. The process is generally carried out using gas proportional counters. While the method is straightforward, it does not distinguish between different alpha-emitting radionuclides. Two primary factors must be considered during this measurement: the reference radionuclide and the burial factor. The first issue is resolved by adopting the concept of global alpha measurement. In France, $^{239}$Pu is commonly used as the reference radionuclide. Figure 1 illustrates the burial effect: in a fibrous filter, aerosols penetrate more deeply into the filter, degrading the energy spectrum, with alpha energies appearing lower. However, when using an PTFE-membrane filter, the alpha spectrum presents almost no tailing, indicating minimal interaction between the alphas and the filter.

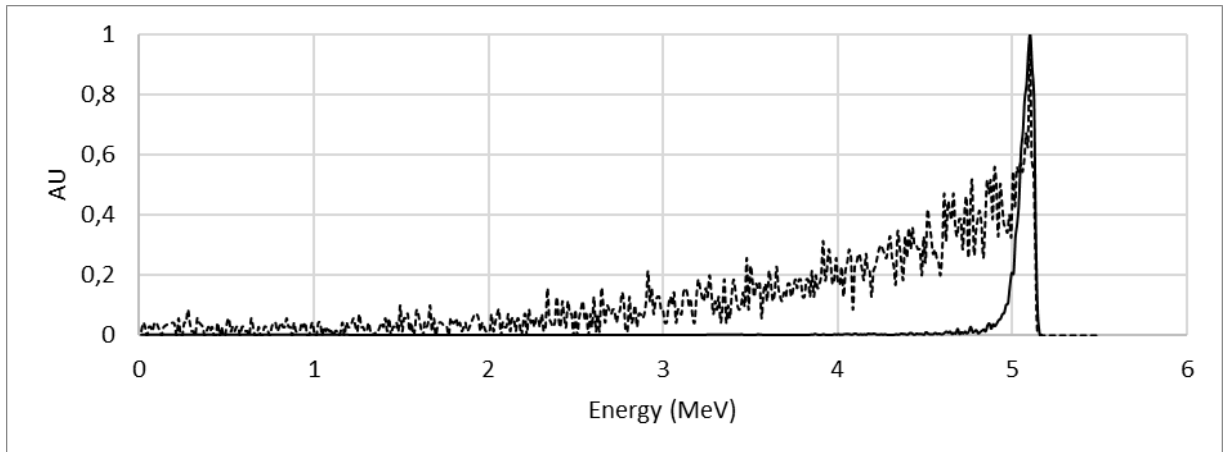

*Figure 1 – Illustration of the burial effect for two filters tagged with $^{239}$Pu. The spectrum from the fibrous filter (dashed line) is significantly broader than the PTFE-membrane filter one (plain line).*

Thus, we propose to modify the reference radioelement to address several of the aforementioned issues. An alternative is $^{241}$Am, a radionuclide that allows non-destructive measurement while preserving traceability to the international system (SI) and maintaining similar properties to $^{239}$Pu for alpha measurements. In view of the characteristics listed in table 1, this is a highly suitable candidate.

*Table 1 – Comparison between $^{239}$Pu and $^{241}$Am (Library for gamma and alpha emissions)*

|  | $^{239}$Pu | $^{241}$Am |
|---|---|---|
| *Main alpha emission and intensity* | 5 156,59 (14) keV – 70,79 (10) %<br>5 143,82 (21) keV – 17,14 (4) %<br>5 105,81 (21) keV – 11,87 (3) % | 5 485,56 (12) keV – 84,45 (10) %<br>5 442,86 (12) keV – 13,23 (10) %<br>5 388,25 (13) keV – 1,66 (3) % |
| *Main gamma emission and intensity* | / | 59,5409 (1) keV – 35,92 (17) %<br>26,3446 (2) keV – 2,31 (8) % |
| *Impurity in liquid standard* | Generally 0.4% of $^{240}$Pu | / |
| *Regulated* | Highly<br>Euratom Treaty (2012/C 327/01)<br>Code de la défense R1333-1 à -19 | / |
| *Standardisation method* | Liquid scintillation counting | Gamma spectrometry |

Given the proximity of the alpha energies (300 keV difference), it can be reasonably assumed that the impact on measurements made using proportional counters will be insignificant. The gamma emission of $^{241}$Am at 59 keV can be accurately measured on GeHPs and is largely unaffected by matrix effects (absorption of energy in the material) for filters. A solution of $^{241}$Am can be produced with a higher level of purity than $^{239}$Pu. Furthermore, americium is more chemically stable than plutonium, thus offering the potential for standard filters to last for longer periods of time. This is particularly advantageous given the target of reusable standard filters with an activity variability of less than 0.5% over a period of one year.

Furthermore, $^{241}$Am is not a nuclear material. Consequently, the implementation of regulations on nuclear material accounting represents a significant advantage in terms of eliminating the presence of $^{239}$Pu in the laboratory.

## 2. Material and Methods

To calibrate detectors, organizations must use reference objects very similar to their routine air sampling filters, marked with a known activity of the reference radionuclide. For this, they rely on IRSN/SCA for manufacturing and the CEA/LNHB for calibration.

The aerodynamic median activity diameter (AMAD) is the most challenging parameter to determine to produce reference filters, as environmental aerosols do not have a well-defined granulometry. Hence, two standard dimensions are generally accepted:

- 0.4 µm for environmental sampling.
- 4 µm for sampling in installations.

The ICARE test bench is a unique facility in Europe, allowing the controlled production of radioactive aerosols (Ammerich, 1989; Monsanglant-Louvet et al., 2015). It is used to produce reference filters by nebulizing a saline solution marked with the desired radionuclide to create an aerosol with the defined AMAD. This atmosphere is sampled through a filter at the specified speed or flow rate until the desired activity is achieved. However, the exact activity cannot be measured with the precision required for calibration at this point. Therefore, the filter is sent to the requesting organization for further calibration.

A total of 42 filters were manufactured on the ICARE test bench specifically for this study. The filters' characteristics are summarized in table 2.

*Table 2 – Summary of the produced filters characteristics*

| Medium | Cellulose fibrous filter | | PTFE membrane filter | |
|---|---|---|---|---|
| Total/Active diameters | 51/47 mm | | 47/37 mm | |
| Aerosol Type | CsCl – KCl | | | |
| AMAD | 4 µm | 0.4 µm | 4 µm | 0.4 µm |
| Filtration flow rate | 20 – 100 L·min$^{-1}$ | | 20 L·min$^{-1}$ | |
| Filtration speed | 0.2 – 1 m·s$^{-1}$ | | 0.4 m·s$^{-1}$ | |
| Nuclide | $^{241}$Am – $^{239}$Pu | | | |
| Activity | 0.1 – 100 Bq | | | |

The filters were produced using a saline solution marked with the target radionuclide. The aerosol production system allowed precise control of the aerodynamic median activity diameter (AMAD). Two types of filters were used: fibrous filter and PTFE membrane filter, with different characteristics, as shown in the table.

To determine the activity of $^{241}$Am, the CEA/LNHB calibrates high-purity germanium (GeHP) detectors with various primary standards across multiple geometries and measurement distances. These standards correspond to 22 radionuclides emitting gamma radiation with energies ranging from 15 keV to 2000 keV.

This extensive calibration process allows the creation of efficiency curves that characterize the detector. The uncertainties in measured efficiency range from 0.4% to 0.6% across the entire energy spectrum. The calibration process spans one year of measurements and calculations, yielding highly precise efficiency values and, consequently, activity measurements.

Finaly, measuring a filter marked with 10 Bq of $^{241}$Am, with a diameter of 47/37 mm, takes one week to achieve a relative uncertainty of 1.5% at k=2.

## 3. Results

Figure 2 presents the obtained granulometries, measured with a DLPI (Dekati Low Pressure Impactor) for the aerosol generators on the ICARE test bench, allowing the production of aerosols with:

- AMAD of 0.386 ± 0.020 μm with a geometric standard deviation of 1.585 ± 0.076;
- AMAD of 4.28 ± 0.18 μm with a geometric standard deviation of 1.376 ± 0.058.

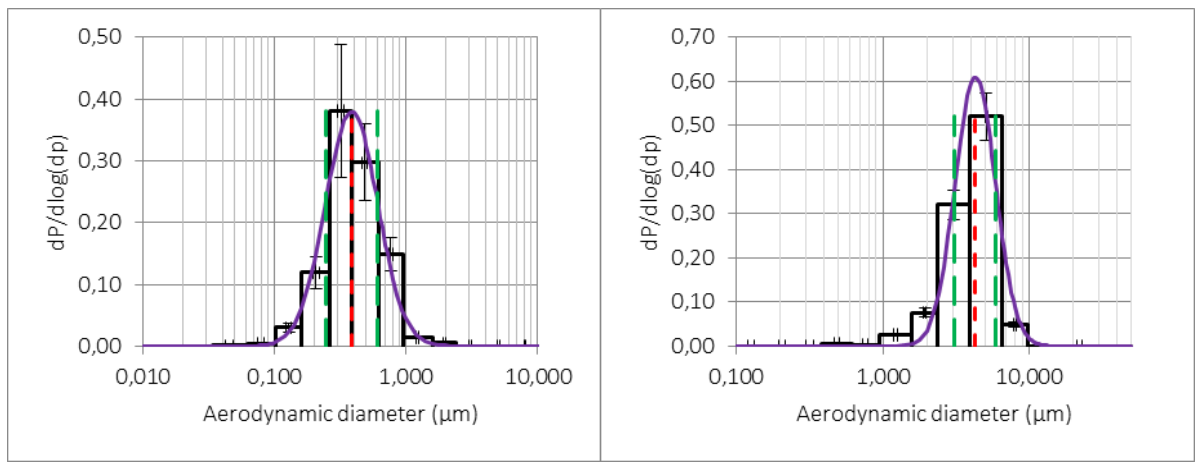

*Figure 2 – Granulometry of produced aerosols on the ICARE test bench for this study. The AMAD of the left figure is 0.386 μm, and 4.28 μm on the right.*

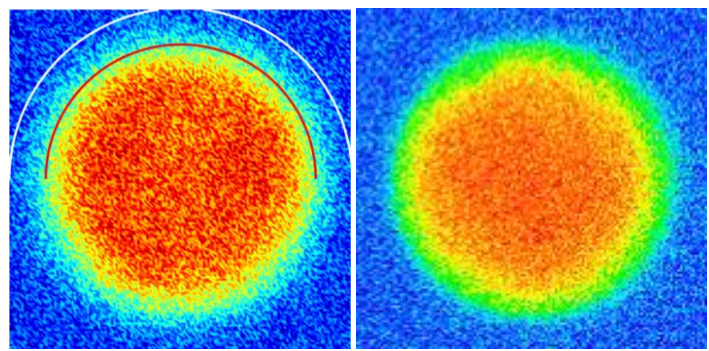

*Figure 3 – On the right, an autoradiography of PTFE-filter tagged au CsCl($^{241}$Am), AMAD = 0.4 μm. On the left side a simulation of this autoradiography, with the active diameter (red) and the filter diameter (white)*

The PTFE-membrane filters allow a consistent calibration between the two radionuclides $^{241}$Am and $^{239}$Pu when using proportional counters. The narrow peaks provided by these filters ensure that the detector's energy window is unaffected by the 300 keV difference between the radionuclides' alpha emissions. The calibration for all the PTFE-membranous filter yields to these efficiencies on our lab's gas proportional counters:

- $^{239}$Pu : 0.360 ± 0.016;
- $^{241}$Am : 0.361 ± 0.008.

These efficiencies are equal; thus, the nuclide change does not affect the calibration.

However, for fibrous filters, the yields for $^{239}$Pu and $^{241}$Am are no longer equal due to the broader peaks exceeding the detector's energy window. An additional correction factor could be determined, but it is preferable to change the reference radionuclide. As expected, the efficiency for $^{241}$Am is lower at 0.260 ± 0.024, and the results are more dispersed.

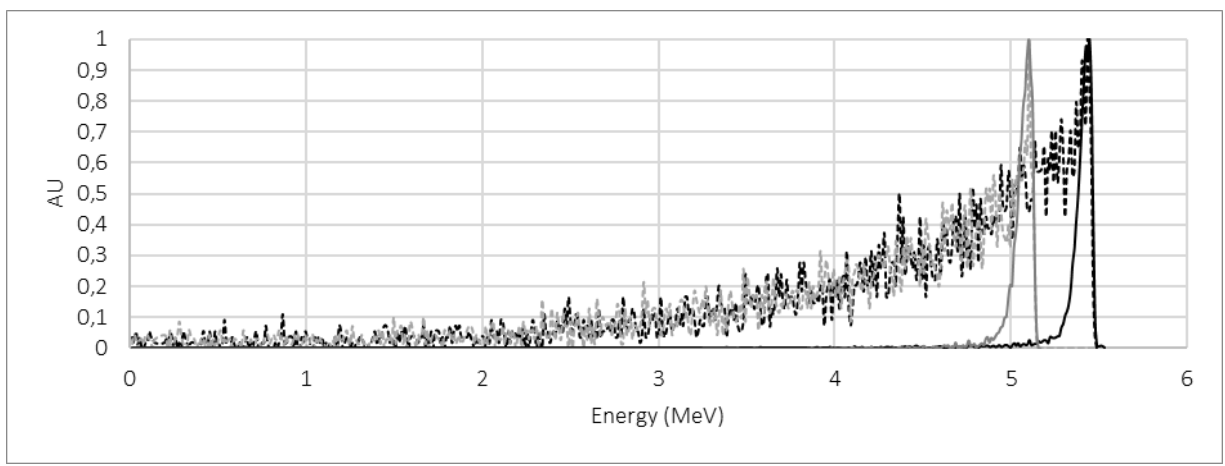

*Figure 4 – Alpha spectrometry of PTFE-membrane filter (plain lines) and fibrous filters (dashed lines) filters for both radionuclides ($^{239}$Pu in grey and $^{241}$Am in black)*

This figure illustrates the challenges posed by fibre filters, particularly the broader peaks, which impact calibration accuracy. This method authorises non-destructive measurement of $^{241}$Am-marked filters, preserving the reference filters for further use.

## 4. Conclusion

This study successfully demonstrates the potential of $^{241}$Am as an alternative to $^{239}$Pu for the non-destructive calibration of filters used in radioactive aerosol measurements. The benefits of using $^{241}$Am include its combined alpha and gamma emissions, higher chemical stability, and reduce regulatory constraints, making it a more versatile and safer option for long-term use. The ICARE test bench enabled the precise production of aerosols with well-controlled granulometries, ensuring accurate calibration.

By utilizing $^{241}$Am, we not only enhance the accuracy and reliability of atmospheric contamination monitoring but also eliminate the need for destructive processes associated with $^{239}$Pu. This innovation represents a significant step forward in radioprotection practices, offering laboratories a robust, efficient and compliant solution for measuring radioactive contamination in air samples. The non-destructive nature of the calibration process, coupled with high precision, opens new possibilities for the reusability of reference filters, reducing costs and improving operational efficiency.

# Acknowledgements

## Funding

No specific funding was provided for this research.

## Conflicts of interest

The authors declare that they have no conflicts of interest.

## Data availability declaration

The research data are included in the article.

## Author contributions

G. Dougniaux: sample fabrication, aerosol study

B. Sabot: sample measurement, gamma spectrometry

S. Pierre: sample measurement, alpha spectrometry

B. Dhieux Lestaevel: sample fabrication

## Ethical approval

Ethical approval was not required.

## Informed consent statement

This article does not contain any studies involving human subjects.